\documentstyle[psfig]{caosp}

\def\Teff{effective temperature} 
 
\def\fq{frequency}
\def\fqs{frequencies}
\def\acfq{acoustic cut-off frequency}
\def\atm{atmosphere}
\def\atms{atmospheres}
\def\md{model}
\def\mds{models}
\def\Ku{Kurucz} 

\def\teff{$T_{{\rm eff}}$}

\def\nuac{$\nu_{\rm M}$}

\def\nuoM{$\nu_{\rm M}^{\rm (1)}$}

\def\nutwM{$\nu_{\rm M}^{\rm (2)}$}

\def\nuthM{$\nu_{\rm M}^{\rm (3)}$}

\begin{document}
\pubyear{1998}
\volume{27}
\firstpage{304}
\htitle{Atmospheric structure and acoustic cut-off frequency of
        roAp stars}
\hauthor{N. Audard {\it et al.}}
\title{Atmospheric structure and acoustic cut-off frequency of 
	roAp stars}   
\author{N.\,Audard \inst{1,2,3}, 
   F.\,Kupka \inst{2}, P.\,Morel \inst{3},  J. Provost \inst{3}  
   \and W.\,W.\,Weiss \inst{2}}

\institute{Institute of Astronomy, University of  Cambridge, Madingley Road, 
Cambridge CB3 OHA, England \and  Institute for Astronomy, University of Vienna, 
T\"urkenschanzstra\ss e 
17, A-1180  Vienna, Austria  \and D\'epartement Cassini, UMR 6529, 
Observatoire de la C\^ote d'Azur, BP 4229, F-06304 Nice cedex 4, 
France }
             
\maketitle

\begin{abstract}
Some of the  rapidly oscillating (CP2) stars, have \fqs\ which are larger
than the theoretical \acfq. 
As the cut-off frequency depends on the $T(\tau )$ relation in the 
atmosphere, we have computed \mds\ and adiabatic \fqs\ for 
pulsating Ap stars with $T(\tau)$ laws based on Kurucz 
model atmospheres and on Hopf's purely radiative relation. 
 
The fre\-quen\-cy-de\-pen\-dent treat\-ment  
of ra\-dia\-ti\-ve trans\-fer as well as an improved calculation 
of the radiative pressure in Kurucz model atmospheres  
increase the theoretical \acfq\  by about 200\,$\mu$Hz, 
which is closer to the observations. 

For $\alpha$\,Cir we find models with Kurucz atmospheres which have 
indeed a cut-off frequency 
beyond the largest observed frequency  and which are well within the \teff\ --
$L$ error box. For HD\,24712 only models which 
are hotter by about 100\,K and less luminous by nearly 10\% than what is 
actually the most probable value would have an \acfq\ large enough. 

One may thus speculate that the old controversy about a 
mismatch between observed largest frequencies
and theoretical cut-off frequencies of roAp star models is resolved.
However, the observational errors for the astrophysical 
fundamental parameters have to be reduced further and the model 
atmospheres refined.

Further details can be found in Audard et al. (1997)

\thesaurus{06(08.01.3, 08.03.2, 08.09.2, 08.15.1, 08.22.3)}

\end{abstract}

     \section{Introduction}
It has been argued by Shibahashi and Saio (1985) that the  
cut-off \fq\ is largely influenced by the $T(\tau)$ relation which
requires a careful modelling of these layers.  
Frequently, atmospheres in stellar \mds\ are based on an
Eddington or Hopf law (e.g. Mihalas 1978), where 
the  radiative transfer is considered 
to be \fq\ independent (grey case), and 
convection is not included. 

We used the 
LTE Kurucz {\sc atlas}9 code (Kurucz, 1993) 
without the ``overshooting option''
(Castelli 1996) to calculate an interpolation table for 
$T(\tau, T_{{\rm eff}})$.  
Model \atms\ with solar 
composition were computed for $\log g = 4.2$ and for \teff\ ranging from 
7400 to 10000\,K, and no
additional contribution to line opacity by microturbulence 
has been assumed. 

The internal structure models  of 1.8\,$M_{\odot}$ representative 
for CP2 stars
were computed with the CESAM code  (Morel 1993 and 1997).
We do not include effects from a magnetic field. 
  
We shall call ``Hopf'' and  ``Kurucz'' models  
full stellar models whose atmospheres are derived  
from  Hopf's law and Kurucz model \atm s respectively.

	\section{Cut-off frequencies}
	\subsection{Theoretical background}		\label{Sfrequ}
To calculate the cut-off frequencies 
we make the  assumptions that 
for low-degree modes  the displacement is essentially vertical,
so that the horizontal component can be neglected.
Consequently, we consider only radial modes and,    
because we investigate modes of high radial orders, 
we adopt the 
Cowling approximation.
We do not include the effects of a magnetic field. 

The {\em cut-off \fq,} \nuac,  is the maximum value 
of the \fq\ above which modes are not at all reflected 
towards the stellar interior  but propagate outwards in the atmosphere. 
We have calculated the cut-off \fq\  according to 
three expressions for the acoustic potential: 
the formulation by Vorontsov \& Zarkhov (1989) (\nuoM), Gough (1986) 
(\nutwM),  and the approximation of an isothermal \atm\ (\nuthM) 
(see e.g.  Gough, 1986).   

Table\,1 gives the values of these cut-off \fqs\ 
for the Hopf and Kurucz \mds\ of 1.8 $M_{\odot}$ and age 500 Myr.
Along the main sequence, the relative difference between 
the cut-off frequencies from  Kurucz and Hopf models of 1.8 M$_{\odot}$ 
models remains almost unchanged and is about 8.5\% (see Tab.\,1)  
(see also Shibahashi, 1991). 

This result is in agreement with results  
of Shibahashi \& Saio (1985) and of Matthews et al. (1990, 1996). 

        \begin{table}[t]
        \small
        \begin{center}
        \caption{Acoustic cut-off frequencies 
		\nuoM, \nutwM\ and \nuthM\ in $\mu$Hz, 
        	for the Hopf and \Ku\ \mds. 
                $\Delta \nu $ is the difference 
                between the respective Kurucz and Hopf models. }  
	\label{t1}
        \begin{tabular}{rcccccc}
        \hline\hline     
        & Hopf   & Kurucz & $\Delta \nu $ \\
\hline	
\nuoM   &  2324  & 2528 & 204\\
\nutwM  &  2354  & 2743 & 389\\
\nuthM  &  2232  & 2465 & 233\\
\hline\hline
	\end{tabular}
	\end{center}
        \end{table}
        \vspace{-2mm}

	\subsection{Comparison with observations}	\label{Sdiscuss}
For about 5 out of 28 known roAp stars, the largest published \fq\
exceeds the expected theoretical
\acfq\ determined from  stellar models whose atmospheres 
are derived from an Eddington or Hopf law. 
We do not consider here roAp stars for which the highest observed 
frequency probably is a harmonic of their nonlinear oscillation. 

Since the uncertainty in the age determination can introduce a 
significant error
in the computed \acfq\ , we will focus our discussion  
on those two roAp stars, HD\,24712 and HD\,128898, for which we
have more reliable mass estimates due to the availability of
{\sc hipparcos} parallaxes. 
The photometric and spectroscopic properties of the roAp star 
{\bf HD\,24712}    
can be reproduced with a model of 
1.63\,$M_{\odot}$, $Z=0.02$ and an age of about 900\,Myr. 
The cut-off \fq\ \nuoM\ is 2280\,$\mu$Hz and 2480\,$\mu$Hz 
for the Hopf and Kurucz models respectively. 

We have also computed Hopf and Kurucz \mds\ with appropriate age for 
{\bf HD\,128898} ($\alpha$\,Cir). A Kurucz 
\md\ \atm\ was calculated with an opacity distribution function  
specific to the composition of $\alpha$\,Cir (Piskunov \& Kupka 1997).  
Stellar models with 1.93\,$M_{\odot}$, 
$Z\,=\,0.03$ and an age of 400\,Myr, fit the observed 
values (Kupka et al. 1996). 
The cut-off frequencies \nuoM\ for the Hopf and Kurucz models
are 2346  and 2600\,$\mu$Hz, respectively.
The \acfq\ computed for the Kurucz model is compatible with the largest
observed frequency.  
However, the cut-off frequency depends on the model input parameters 
and one must 
therefore account for uncertainties inherent to observations and modelling. 
Evolutionary tracks and lines of constant cut-off frequency for Kurucz 
models are plotted in Fig.\,1a together with the 
observational error boxes. 

Asteroseismology is a powerful tool for 
determining the evolutionary status of stars via the 
frequency separation 
$\nu_0 = (2~{\int_0^R(dr/c)})^{-1}  
\sim \nu_{n, \ell} - \nu_{n-1, \ell}$ 
(see e.g. Shibahashi 1991, and Kurtz \& Martinez 1993).   
If $\nu_0$ can be measured, a more reliable 
estimate of the evolutionary status can be derived than with 
our classical approach.
Fig.\,1b  shows lines of constant frequency spacing 
$\nu_0$ for the same models as for Fig.\,1a. 
The observed value $\nu_0=68\,\mu$Hz for HD 24712 (Kurtz et al. 1989) 
is consistent with our classically determined error box, 
while there is a serious problem for $\alpha$ Cir, because 
the observed value of $\nu_0$ is $50\,\mu$Hz (Kurtz et al. 1994) 
which cannot be reconciled with the spectroscopically determined \Teff\ and/or
the luminosity.  

\begin{figure}
\centerline{
\psfig{figure=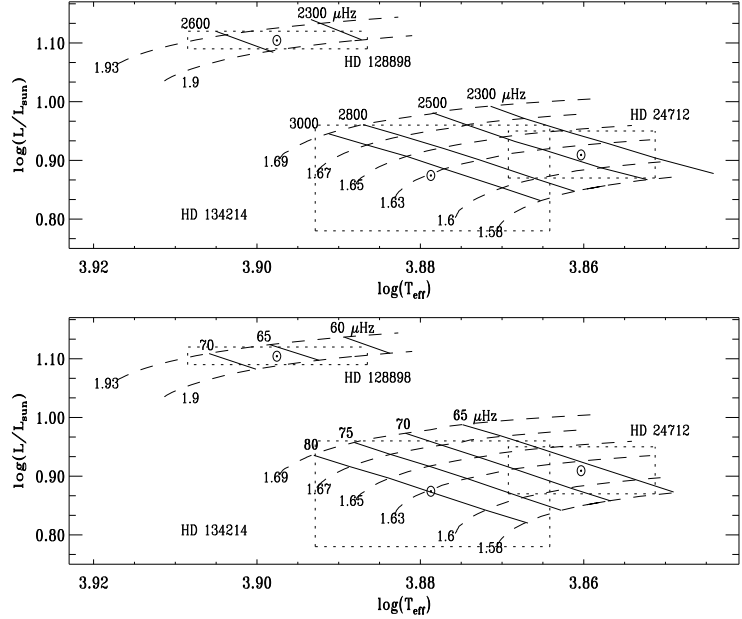,height=8.8cm}}
              \caption{HR diagram for stars with 1.58 $M_{\odot}$ 
                to 1.69 $M_{\odot}$ for $Z=0.02$ 
                (age up to 1000 Myr), and with 1.90 and 1.93 
                $M_{\odot}$ for $Z=0.03$ (age up to 700 Myr) 
                (dashed lines).  The roAp
                stars HD\,24712, HD\,134214 and HD\,128898 are indicated 
                by circles and error boxes. 
                Full lines are lines of  constant cut-off frequency  
                $\nu_{\rm M}^{(1)}$  for the Kurucz models, for
                2300, 2500, 2800 and 3000\,$\mu$Hz for $Z=0.02$, 
                and for 2300 and 2600\,$\mu$Hz for $Z=0.03$  
                ({\bf a}), and lines of constant frequency 
                spacing $\nu_0 = (2~{\int_0^R(dr/c)})^{-1}$ 
                for the same models, from 80  
                to 65\,$\mu$Hz for  $Z=0.02$, and from 70 to 
                60\,$\mu$Hz for  $Z=0.03$. For HD\,24712
            the frequency splitting $\nu_0$ = 68\,$\mu$Hz (Kurtz et al. 1989), 
		and 50\,$\mu$Hz (Kurtz et al. 1994)
		for HD\,128898 ({\bf b}).}
        \end{figure}
	\vspace{-3mm}
	\section{Conclusion}
        \vspace{-1mm}
We  have shown that along 
the main sequence, Kurucz model atmospheres increase the cut-off \fq\ by 
about 8.5\,\% relative to the value derived from the Hopf $T(\tau)$ relation. 

For  HD\,24712 and $\alpha$\,Cir, 
we find models with Kurucz atmospheres and with parameters in agreement with 
the observational error box which have a theoretical cut-off frequency 
larger than the largest observed \fq\ and hence are in agreement with 
observations. One may thus speculate that the old controversy about 
a mismatch between observed largest frequencies
and theoretical cut-off frequencies of roAp star models is resolved.

However, effects from e.g. an abundance different from solar 
might affect the cut-off frequency.  
Abundant rare-earth elements, through blanketing effects,
could decrease the surface temperature and thus increase the cut-off \fq. 
This \fq\ might also be affected by a chemical composition gradient
(Vauclair \& Dolez 1990). Magnetic field (Dziembowski \& Goode 1996) 
and NLTE effects should also be taken into account.
\\

       \acknowledgements 
We are indebted to Maurice Gabriel and  Hiromoto Shibahashi 
for their comments and suggestions. 
Computing resources
and financial support for this international collaboration
were provided by the Fonds zur F\"orderung
der wissenschaftlichen  Forschung project {\em S 7303-AST}
and Digital
Equipment Corp. (Europe External Research Program,
project {\em STARPULS}). NA is grateful to PPARC  (UK) 
for financial support. 

\end{document}